\documentclass[aps,prd,preprint,groupedaddress]{revtex4}
\usepackage{epsf,epsfig}
\newcommand{\Exp}[1]{{\mathrm{e}}^{#1}}
\newcommand{\be}{\begin{equation}}
\newcommand{\ee}{\end{equation}}
\newcommand{\bea}{\begin{eqnarray}}
\newcommand{\eea}{\end{eqnarray}}
\newcommand{\LL}{{\mathcal L}}
\newcommand{\pa}{\partial}
\newcommand{\pam}{{\partial_\mu}}

\newcommand{\ba}{\begin{array}}
\newcommand{\ea}{\end{array}}
\newcommand{\lp}{\left(}
\newcommand{\rp}{\right)}
\newcommand{\Bv}{\displaystyle{\biggl\vert}}
\newcommand{\lac}{\left\{}

\renewcommand{\a}{\alpha}

\newcommand{\g}{\gamma}
\renewcommand{\d}{\delta}

\newcommand{\om}{\omega}

\newcommand{\ep}{\epsilon}

\begin{document}

                
\title{Q-Ball Condensation}


\author{Stephen S. Clark}
\email{stephen.clark@a3.epfl.ch}
\affiliation{CEC M. de Stael,\\ 25 route de st Julien,\\ 1227 Carouge, Switzerland.}


\date{\today}

\begin{abstract}
Q-balls arise in particle theories with $U(1)$ global symmetry. The coupling of the corresponding scalar field to fermions leads to Q-ball evaporation. 
In this paper we consider the oposite problem, the case where a Q-ball absorbs particles to grow. In particular we shall use
the exact quantum mechanical description of fermions interacting with a Q-ball to solve the problem. Results show that Q-ball condensation can be 
another mechanism for Q-ball creation.

\end{abstract}


\maketitle

\section{Introduction}
A scalar field theory with an unbroken continuous global symmetry admits a remarkable class of solutions, non-topological solitons or
Q-Balls. These solutions are spherically symmetric non-dissipative solutions to the classical field equations \cite{Qballscoleman,Qballscohen,Qballsnew}. 
In a certain way they can be viewed as a sort of Bose-Einstein condensate of ``classical'' scalars. 
The construction of these solutions is made by building a ground state of energy in the sector of fixed charge.
An important amount of work has been done on Q-Ball dynamics and on their stability versus decay into scalars \cite{Qballscoleman,axenides} :  
the stability of Q-Balls is due to the fact that their mass is smaller than the mass of a collection of scalars. 

Some of the standard Q-ball properties, mass proportional to $Q^{3/4}$, size proportional to $Q^{1/4}$ makes them good candidates for self
interacting dark matter \cite{Kusenko:1997si,Kusenko:2004yw,Kusenko:2001vu}. For Q-balls to be good candidates for dark matter they need to
have a very big charge, $Q>10^{15}$, this enormous size ensures also the stability of Q-balls.

The most popular model allowing ``big'' scalar field is the Affleck-Dine Baryogenesis process \cite{Aff, Paw, Mul1}. In this picture a large charged scalar
condensate will break up into smaller objects that have Q ball properties. This scalar condensate is the one required for Affleck-Dine baryogenesis.
This scalar field carries baryon (or lepton) number and has a very flat potential so it can be given a large expectation value. 
This breaking up is done through supersymmetry breakdown. We have two major ways for breaking up supersymmetry, 
the gauge or gravity mediated mechanism. Both mechanisms will lead to different type of Q-balls. 
The major difference is that the Q ball's size might not depend on its charge, while its charge will be linked to the one of the initial scalar field.
One other way we could create Q balls is by solitosynthesis, a process of charge accretion around a Q ball seed \cite{axenides,Post}. 
This model has the advantage not to need any complicated symmetry breaking.

All these creation processes use only the scalar field itself or small Q-balls to create bigger Q-ball. The method we shall use to create big Q-balls is
different. Once we have the exact quantum mechanical description of an interacting Q-ball we shall build a state where no fermions move away from the
Q-ball. This construction ensures  that we do not deal with scattering of a Q-ball. This state we can build explicitly describes a Q-ball absorbing fermions
to grow. This alternative construction provides a much simpler way to obtain big Q-balls.
First we shall give a brief overview of Q-balls and their properties to then describe particle absorption before concluding.
\section{Q-Balls}
We review here the basic properties of a 3-dimensional Q-Ball using the simplest possible model. As we mentioned in the introduction, the Q-Ball is the
ground state of a scalar theory containing a global symmetry. 
We can now build the simplest model in $3\oplus1$ dimensions having a Q-Ball solution: it is a $SO(2)$ invariant theory 
of two real scalar fields (in fact it is the $U(1)$ theory of one complex scalar field) \cite{Qballscoleman}. We start by writing down the Lagrangian and the
equations of motion for the scalar field, to obtain the conserved charge and current.
The Lagrangian of the scalar sector is given by :
\begin{eqnarray}
\LL=\pam\phi^\star\pa^\mu\phi-U(|\phi|).
\end{eqnarray}
The $U(1)$ symmetry is
\begin{eqnarray}
\phi\rightarrow\Exp{i\alpha}\phi. \nonumber
\end{eqnarray}
The conserved current is 
\begin{eqnarray}
j_\mu=i(\phi^\star\pam\phi-(\pam\phi^\star)\phi),
\end{eqnarray}
and the conserved charge is
\begin{eqnarray}
Q=\int d^3xj_0.
\end{eqnarray}
It was shown in \cite{Qballscoleman} that new particles (Q-Balls) appear in the spectrum, if the potential is such that the 
minimum of $2\frac{U(|\phi|)}{|\phi|^2}$ is at some value 
$\phi_0\neq0$.
\begin{eqnarray}
Min[2U/|\phi|^2]=2U(\phi_0)/|\phi_0|^2<\mu^2=U''(0).
\end{eqnarray}
The charge and energy of a given $\phi$ field configuration are :
\begin{eqnarray}
\ba{c}Q=\int(-\pa_t\phi^\star\phi+c.c.)d^3x,  \\
E=\int\left[\frac{1}{2}|\dot{\phi}|^2+\frac{1}{2}|\nabla\phi|^2+U(\phi)\right]d^3x.\ea\label{Qdef}
\end{eqnarray}
The Q-Ball solution is a solution with minimum energy for a fixed charge, we thus introduce the following Lagrange
multiplier
\begin{eqnarray}
{\mathcal E}_\om=E+\om[Q-\int(\phi^\star\pa_t\phi+c.c.)d^3x].
\end{eqnarray}
Minimising this functional with the standard  Q-Ball ansatz :
\begin{eqnarray}
\phi=\phi(\vec{x})\Exp{i\om t},
\end{eqnarray}
where $\phi(r)$ is a monotonically decreasing function of distance to the origin, and zero at infinity, will give all Q-ball properties and relations linking $\om_0$,
$\phi_0$. This $\phi$ function admits a lot of solutions \cite{Qballsnew}, but we shall use for simplicity the step function profile.
\section{Absorption of massless fermions}

To solve the problem of particle absorptions, we s first solve the equations of motion and obtain the Heisenberg
field operator representing a fermion interacting with a Q-ball \cite{clark1}. In one space dimension this solution will be expressed in the form,
\begin{eqnarray}
\Psi_Q=\frac{1}{\sqrt{4\pi}}\int d\ep\lp\psi_Q^+(\ep,t,z)A(\ep)+\psi_Q^-(\ep,t,z)B(\ep)\rp, \nonumber
\end{eqnarray}
where the $\psi^{\pm}_Q(\ep,t,z)$ are a basis of the solution to the Dirac equation for fermions interacting with a Q-ball of charge Q.
$A(\ep)$ and $B(\ep)$ are operators depending on energy, their anti-commutation relations are the standard ones if the $\psi$
solutions satisfy proper orthogonality conditions.
The next step we shall use is consider the space asymptotics of this solution. Far away from the Q-ball
($z=\pm\infty$ for one space dimension) the solution is the standard free field solution. This identification will give us
a relation between the solution operators $A(\ep)$, $B(\ep)$ and the free asymptotic ones $a(p)$, $b(p)$. The only
difficulty in this identification is that the quantisation of the solution was made using energy (due to the time dependence of interaction)
while the asymptotical operators depend on momentum. The next step will be to define and solve the particle absorption condition, saying that no particles
are coming out of the Q-ball. In terms of asymptotic operators it is 
\begin{eqnarray}
\ba{c}a_L(p)|\Psi>=b_L(p)|\Psi>=0\quad\mbox{for $p<0$, on the left} \\
a_R(p)|\Psi>=b_R(p)|\Psi>=0\quad\mbox{for $p>0$, on the right.}\ea
\end{eqnarray}
The last step of the resolution will consist in using the total Heisenberg operator $\Psi$, and the particle absorption state to compute the
fermionic flux giving condensation rate. The main idea here is to use a solution containing only (anti-)particles going inside the Q-ball and never coming 
out again.

\subsection{Solutions to the equations of motion}
Writing down the Lagrangian of a massless fermion having a Yukawa interaction with a scalar field (the Q-ball) gives, 
\begin{eqnarray}
\LL_{ferm.}=i\bar{\psi}\sigma^\mu\pam\psi+(g\phi\bar{\psi}^C\psi+h.c),
\end{eqnarray}
where the $C$ superscript indicates the charge conjugated fermion. 
The equations of motion and their solutions are fully described in literature on the subject (\cite{Qballscoleman,Qballscohen,Qballsnew}). For simplicity we shall
consider the $1\oplus1$ dimensional case, in this case we use a Majorana representation of the $\gamma$-matrices, that is :
\begin{eqnarray}
\gamma^0=\sigma^1\quad\mbox{and}\quad\gamma^1=i\sigma^2,\nonumber
\end{eqnarray}
and the charge conjugation given by
\begin{eqnarray}
\psi^C=\sigma^3\psi^\star.\nonumber
\end{eqnarray}
Instead of treating separately the fermion
and the anti-fermion, we shall construct the exact global solution to this problem, this solution will be made of different parts first 
the solution inside the Q-Ball (for $z\in[-l,l]$) and the solution outside the Q-ball that we shall then match togther.
The equations of motion for $\Psi=\lp\ba{c}\psi_1 \\ \psi_2\ea\rp$  are : 
\begin{eqnarray}
\ba{c}(i\pa_0+i\pa_z)\psi_1-g\phi\psi_2^{\star}=0, \\
(i\pa_0-i\pa_z)\psi_2^{\star}-g\phi^\star\psi_1=0. \ea
\end{eqnarray}
and $\phi=\phi_0\Exp{-i\om_0t}$ in the zone from $-l$ to $+l$ and zero everywhere else and $g$ real. We call $z$ the only spatial dimension.

The solution to this system is \cite{clark1} :
\begin{eqnarray}
\Psi_Q=\frac{1}{\sqrt{4\pi}}\int d\ep\Exp{-i\ep t}\lp\psi_Q^+(\ep,z)A(\ep)+\psi_Q^-(\ep,z)B(\ep)\rp
\Exp{i\frac{\om_0}{2}z}\Omega(t), \label{solQBall} 
\end{eqnarray}
with
\begin{eqnarray}
\Omega(t)=\lp\ba{cc}\Exp{-i\frac{\om_0}{2}t} & 0 \\ 0 & \Exp{i\frac{\om_0}{2}t} \ea\rp,
\end{eqnarray}
{\small
\begin{eqnarray}
\psi^\pm(\ep,z)=\lp\ba{c} \lp\ba{c}f_1^\pm(\ep,l)\Exp{i\ep z} \\
(f_2^\pm(\ep,l))^\star\Exp{-i\ep z}\ea\rp z<-l \\ \\
\frac{1}{\sqrt{N_\pm}}\lp\ba{c}(\pm\Exp{-ik_\ep z}+\a_\ep\Exp{ik_\ep z})
\\ (\pm\a_\ep\Exp{-ik_\ep z}+\Exp{ik_\ep z}) \ea\rp -l\leq z\leq +l \\
\\ \lp\ba{c} f_1^\pm(\ep,-l)\Exp{i\ep z} \\ (f_2^\pm(\ep,-l))^\star\Exp{-i\ep
z}\ea\rp z>+l \ea \rp, \label{solution3}
\end{eqnarray}}
the functions $f^\pm_{1,2}$  having the form 
\begin{eqnarray}
\ba{c}f_1^\pm(\ep,l)=\frac{1}{\sqrt{4\pi N_\pm}}\Exp{i\ep l}(\pm\Exp{ik_\ep
l}+\a_\ep\Exp{-ik_\ep l}), \\ 
f_2^\pm(\ep,l)=\frac{1}{\sqrt{4\pi N_\pm}}\Exp{i\ep
l}(\pm\a^\star_\ep\Exp{-ik^\star_\ep l}+\Exp{ik_\ep l}).\ea
\end{eqnarray}
\begin{eqnarray}
N_\pm&=&4\pi\lp\cosh[\mathrm{Im}[k_\ep] l](1+|\a_{\ep}|^2)\right. \nonumber \\
&\pm&\left.\cos[\mathrm{Re}[k_\ep]l]\mathrm{Re}[\a_\ep]\rp,
\end{eqnarray}
and
\begin{eqnarray}
\alpha_\ep=\frac{\ep+k_\ep}{M},\quad k_\ep=\sqrt{\ep^2-M^2} \\
M=g\phi_0.
\end{eqnarray}
Finally the time-dependent matrix was introduced for simplicity, the $N_{\pm}$ are the normalisation constants. Note that $\phi_0$ and $\om_0$ are the 
parameters describing the Q-ball so they both depend on the Q-ball charge $Q$ (it is a fermionic or barionic charge).
Quantisation of solution (\ref{solQBall}) is done using equal time anti-commutation relations for $\Psi$. Since the $\psi^\pm$ functions satisfy 
$\int dz(\psi^{\sigma'}(\ep'))^\dagger\psi^\sigma(\ep)=\delta_{\sigma'\sigma}\delta(\ep'-\ep)$
we can show that,
\begin{eqnarray}
\{A(\ep),A^\dagger(\ep')\}&=&\int dz\int dz'{\lp\psi_Q^+(z,\ep)\rp}^\dagger{\lp\psi_Q^+(z',\ep')\rp}
\times\underbrace{\{\hat{\Psi}_Q,{\lp\hat{\Psi}_Q'\rp}^\dagger\}}_{\d(z'-z)}\nonumber \\
&=&\d(\ep'-\ep) \nonumber \\
&=&\{B(\ep),B^\dagger(\ep')\}
\end{eqnarray}
The $\Psi_Q$ solution we obtained has now being upgraded to a Heisenberg field operator describing fermions interacting with a Q-ball.
\subsection{Relation to asymptotic operators}
Since the Q-ball is localised in space the solution can be considered free far away from the Q-ball. We can thus obtain a relation between
free operators and our interacting ones (see \cite{clark1,abdalla} for details). These relations are,
\begin{eqnarray}
\frac{1}{\sqrt{2\pi}}a_L(p)&=&[f_1^+(\ep,l)A(\ep)+f_1^-(\ep,l)B(\ep)]\Bv_{\ep=p-\frac{\om_0}{2}}\theta(p)+ \nonumber \\
&+&[f_2^+(\ep,l)A^\dagger(\ep)+f_2^-(\ep,l)B^\dagger(\ep)]\Bv_{\ep=p+\frac{\om_0}{2}}\theta(-p) ,\nonumber \\ \label{gauche1}\\
\frac{1}{\sqrt{2\pi}}b_L^\dagger(-p)&=&[f_1^+(\ep,l)A(\ep)+f_1^-(\ep,l)B(\ep)]\Bv_{\ep=p-\frac{\om_0}{2}}\theta(-p)
+ \nonumber \\
&+&[f_2^+(\ep,l)A^\dagger(\ep)+f_2^-(\ep,l)B^\dagger(\ep)]\Bv_{\ep=p+\frac{\om_0}{2}}\theta(p) \nonumber.
\end{eqnarray}
on the left-hand side and,
\begin{eqnarray}
\frac{1}{\sqrt{2\pi}}a_R(p)&=&[f_1^+(\ep,-l)A(\ep)+f_1^-(\ep,-l)B(\ep)]\Bv_{\ep=p-\frac{\om_0}{2}}\theta(p)+ \nonumber \\
&+&[f_2^+(\ep,-l)A^\dagger(\ep)+f_2^-(\ep,-l)B^\dagger(\ep)]\Bv_{\ep=p+\frac{\om_0}{2}}\theta(-p),\nonumber \\ \label{droite2} \\
\frac{1}{\sqrt{2\pi}}b_R^\dagger(-p)&=&[f_1^+(\ep,-l)A(\ep)+f_1^-(\ep,-l)B(\ep)]\Bv_{\ep=p-\frac{\om_0}{2}}\theta(-p)
+ \nonumber \\
&+&[f_2^+(\ep,-l)A^\dagger(\ep)+f_2^-(\ep,-l)B^\dagger(\ep)]\Bv_{\ep=p+\frac{\om_0}{2}}\theta(p).\nonumber
\end{eqnarray}
on the right-hand side. We can check all the standard properties of these operators so they can be used to define the particle absorbing state.
\subsection{Construction of the particle-absorbing state}
\begin{figure}
\begin{center}
\includegraphics{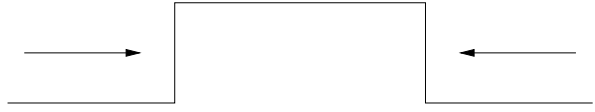}
\caption{Description of particle absorbing state.}
\label{figcurrent1}
\end{center}
\end{figure}
The construction of this quantum state $\Psi$ will be done using the fact that there are no particles leaving
 the Q-Ball. The only particles we want to have are positive momentum particles on the left and negative momentum particles on the right. 
We can express this condition in terms of the free operators as :
\begin{eqnarray}
\ba{c}a_L(p)|\Psi>=b_L(p)|\Psi>=0\quad\mbox{for $p<0$, on the left} \\
a_R(p)|\Psi>=b_R(p)|\Psi>=0\quad\mbox{for $p>0$, on the right.}\ea
\end{eqnarray}
This construction will lead to the opposite sign of the fermionic current on the left and on the right hand side of Q-Ball
using eqs. (\ref{gauche1}-\ref{droite2}). We then obtain four equations. For negative $p$, we have :
\begin{eqnarray}
\ba{c}(f_2^+(\ep,l)A^\dagger(\ep)+f_2^-(\ep,l)B^\dagger(\ep))\Bv_{\ep=p+\frac{\om_0}{2}}|\Psi>=0, \\
(f_2^+(\ep,l))^\star A(\ep)+(f_2^-(\ep,l))^\star B(\ep)\Bv_{\ep=-p+\frac{\om_0}{2}}|\Psi>=0,\ea \label{relp1}
\end{eqnarray} 
and for positive $p$ 
\begin{eqnarray}
\ba{c}(f_1^+(\ep,-l)A(\ep)+f_1^-(\ep,-l)B(\ep))\Bv_{\ep=p-\frac{\om_0}{2}}|\Psi>=0, \\
(f_1^+(\ep,-l))^\star A^\dagger(\ep)+(f_1^-(\ep,-l))^\star B^\dagger(\ep))\Bv_{\ep=-p-\frac{\om_0}{2}}|\Psi>=0.\ea \label{relp2}
\end{eqnarray}
Due to the relation between $\ep$, $p$, $\frac{\om_0}{2}$ given in the subindices of eqs. (\ref{relp1}, \ref{relp2}) 
and the fact that $p$ is either positive or negative, we can 
identify three ranges for $\ep$ :
\begin{itemize}
\item For $\ep<-\frac{\om_0}{2}$ we only have the following two equations :
\begin{eqnarray}
\ba{c}(f_2^+(\ep,l)A^\dagger(\ep)+f_2^-(\ep,l)B^\dagger(\ep))|\Psi>=0, \\
((f_1^+(\ep,-l))^\star A^\dagger(\ep)+(f_1^-(\ep,-l))^\star B^\dagger(\ep))|\Psi>=0.\ea
\end{eqnarray} 
\item $\ep>+\frac{\om_0}{2}$ we have :
\begin{eqnarray}
\ba{c}((f_2^+(\ep,l))^\star A(\ep)+(f_2^-(\ep,l))^\star B(\ep))|\Psi>=0, \\
(f_1^+(\ep,-l)A(\ep)+f_1^-(\ep,-l)B(\ep))|\Psi>=0.\ea
\end{eqnarray}
\item For the middle range $\ep\in[-\frac{\om_0}{2},+\frac{\om_0}{2}]$ we have :
\begin{eqnarray}
\ba{c}(f_2^+(\ep,l)A^\dagger(\ep)+f_2^-(\ep,l)B^\dagger(\ep))|\Psi>=0, \label{un}\\
(f_1^+(\ep,-l)A(\ep)+f_1^-(\ep,-l)B(\ep))|\Psi>=0.\ea \label{deux}
\end{eqnarray}
\end{itemize} 
In the two ranges $\ep<-\frac{\om_0}{2}$ and $\ep>+\frac{\om_0}{2}$ the solution is the trivial one leading to no absorption
\begin{eqnarray}
\ba{c}A(\ep)|\Psi>=B(\ep)|\Psi>=0\quad\mbox{for}\quad\ep>\frac{\om_0}{2}, \\
A^\dagger(\ep)|\Psi>=B^\dagger(\ep)|\Psi>=0\quad\mbox{for}\quad\ep<-\frac{\om_0}{2}.\ea
\end{eqnarray}
In fact these two equations are the same, because we can always use the transformation $A(\ep)=A'(\ep)\theta(\ep)+
{B'}^\dagger(\ep)\theta(-\ep)$, all equations will have vacuum solutions. Here the anti-commutation relations are trivial 
to check because of the two different energy ranges. For the middle range $\ep\in[-\frac{\om_0}{2},+\frac{\om_0}{2}]$
things are a little more complicated, this being the range where particle absorbing occurs. It is also the range of Q-ball evaporation
as first shown in \cite{Qballscohen}. Taking a look at solution (\ref{solution3}) in this range, only
anti-particles are absorbed and changing the sign of $\om_0$ changes the particle type. In fact particles are Majoranna fermions so we identify
particle and anti-particles regarding in which energy range they are. We now need to check normalisation of these
new operators describing the absorbing state and their anti-commutation relations. 
Defining the absorption operators in all the energy ranges, we have   
\begin{eqnarray}
a_a(\ep)=\lac\ba{c}A^\dagger(\ep)\quad\ep<-\frac{\om_0}{2} \\
\sqrt{8\pi}(f_2^+(\ep,l)A^\dagger(\ep)+f_2^-(\ep,l)B^\dagger(\ep))\quad\ep\in[-\frac{\om_0}{2},+\frac{\om_0}{2}] \\
A(\ep)\quad\ep>+\frac{\om_0}{2}\ea\right.\label{aevap} ,
\end{eqnarray}
and
\begin{eqnarray}
b_a(\ep)=\lac\ba{c}B^\dagger(\ep)\quad\ep<-\frac{\om_0}{2} \\
\sqrt{8\pi}(f_2^+(\ep,l)^\star A^\dagger(\ep)-f_2^-(\ep,l)^\star B^\dagger(\ep))\quad\ep\in[-\frac{\om_0}{2},+\frac{\om_0}{2}] \\
B(\ep)\quad\ep>+\frac{\om_0}{2}\ea\right.\label{bevap},
\end{eqnarray}
where the $\sqrt{8\pi}$ factor is the normalisation $\frac{1}{\sqrt{|f_2^+(\ep,l)|^2+|f_2^-(\ep,l)|^2}}$. We also used the fact that 
$f^1_-(\ep,l)=-(f^2_-(\ep,-l))^\star$.
The anti-commutation
relations of these operators are easy to check. They use the fact that $|f_1^\pm|^2=\frac{1}{4\pi}$. The particle absorbing state is now fully
characterised  by the relation :
\begin{eqnarray}
a_a(\ep)|\Psi>=b_a(\ep)|\Psi>=0.
\end{eqnarray}
This simple relation gives the ground state for a Q-Ball absorbing fermions. 
\subsection{Particle absorption rate}
The particle absorption rate is given by the spatial component of current operator  $j^\mu(x)=\bar{\psi}(x)\g^\mu\psi(x)$, which in our case is 
\begin{eqnarray}
\psi_1^\star\psi_1-\psi_2^\star\psi_2=\vec{j}(x)
\end{eqnarray}
that we shall apply on the absorbing state defined by the vacuum for 
$a_a$ and $b_a$ operators. First we invert the systems (\ref{aevap}) and (\ref{bevap}) to obtain :
\begin{eqnarray}
&\bullet&\ep<-\frac{\om_0}{2}\quad \lac\ba{c} A^\dagger(\ep)=a_a(\ep)\\B^\dagger(\ep)=b_a(\ep)\ea\right. \\
&\bullet&\ep>\frac{\om_0}{2}\quad \lac\ba{c} A(\ep)=a_a(\ep)\\B(\ep)=b_a(\ep)\ea\right. \\
&\bullet&\ep\in[-\frac{\om_0}{2},+\frac{\om_0}{2}]\quad \lac\ba{c}A(\ep)=\frac{1}{\sqrt{8\pi}2f_2^+(\ep,l)}(a_c(\ep)^\dagger
+b_a(\ep))\\B(\ep)=\frac{1}{\sqrt{8\pi}2f_2^-(\ep,l)}(a_a^\dagger(\ep)-b_a(\ep))\ea\right.
\end{eqnarray}
Now we can compute the first term of the current on the left hand side of the Q-Ball :
Using anti-commutation relations and the separate range of integrals and the definition of $A(\ep)$ and $B(\ep)$ in
terms of absorption operators $a_e(\ep)$, $b_e(\ep)$ we obtain :
\begin{eqnarray}
<0|\psi_1^\dagger\psi_1|0>&=&\int_{-\infty}^{\frac{\om_0}{2}}d\ep(|f_1^+(\ep,l)|^2<0|a_a(\ep)a_a^\dagger(\ep)|0>
+|f_1^-(\ep,l)|^2<0|b_a(\ep)b_a^\dagger(\ep)|0>) \nonumber \\
&+&\frac{1}{8\pi}\int_{-\frac{\om_0}{2}}^{+\frac{\om_0}{2}}d\ep{\Bv\frac{f_1^+(\ep,l)}{2(f_2^+(\ep,l))^\star}-\frac{f_1^-(\ep,l)}
{2(f_2^+(\ep,l))^\star}\Bv}^2<0|b_e(\ep)b_e^\dagger(\ep)|0>.
\end{eqnarray}
The other term  of the current, proportional to $\psi_2^\star\psi_2$, 
reads :
\begin{eqnarray}
<0|\psi_2^\dagger\psi_2|0>&=&\frac{1}{8\pi}\int_{-\frac{\om_0}{2}}^{+\frac{\om_0}{2}}d\ep{\Bv\frac{(f_2^+(\ep,l))^\star}
{2f_2^+(\ep,l)}
+\frac{(f_2^-(\ep,l))^\star}{2f_2^-(\ep,l)}\Bv}^2<0|a_a(\ep)a_a^\dagger(\ep)|0> \\
&+&\int_{+\frac{\om_0}{2}}^{+\infty}d\ep\left({|(f_2^+(\ep,l))|}^2<0|a_a(\ep)a_a^\dagger(\ep)|0>+
{|(f_2^-(\ep,l))^\star|}^2<0|b_a(\ep)b_a^\dagger(\ep)|0>\right). \nonumber
\end{eqnarray}
A simple calculation shows that terms with infinite bounds will compensate. Leading to :
\begin{eqnarray}
\vec{j}_L&=&\int_{-\frac{\om_0}{2}}^{+\frac{\om_0}{2}}d\ep{\Bv\frac{(f_1^+(\ep,l))^\star}{2f_2^+(\ep,l)}
+\frac{(f_1^-(\ep,l))^\star}{2f_2^-(\ep,l)}\Bv}^2 \nonumber \\
&=&\int_{-\frac{\om_0}{2}}^{+\frac{\om_0}{2}}d\ep\Bv\frac{\a_\ep\sinh[2ik_\ep l]}{\Exp{2ik_\ep l}-\a_\ep^2
\Exp{-2ik_\ep l}}\Bv^2\label{current}.
\end{eqnarray}
This expression is in fact a ``charge modification rate'' $\frac{dQ}{dt}$, we can show it directly using charge conservation (or even the definition of current).
This expression is the same as the one for the evaporation rate \cite{clark1}, even if this result is very intuitive, if a Q-ball can produce particles he can
absorb anti-particles. It gives a new construction to create big Q-balls.
\section{Conclusions}
In this work we showed a alternative model of Q-ball creation, this model is based on Q-ball condensation. Instead of producing fermions the condensing
Q-ball will absorb anti-fermions. Once we had the exact quantum mechanical description of fermions interacting with Q-balls we can impose that no particles 
are coming from the Q-ball. This construction leads to no scattering and therefor only particle absorption. This construction leads to a new way of 
creating Q-balls.

We also showed that absorption rate is the same as evaporation rate, this is is fact the reason why we used  Q-ball condensation to describe this
phenomenon. This result has new implications in the way we understand Q-ball creation and Q-ball life. One of them could be : Let's imagine a small Q-ball
created through the Affleck-Dine process, the Q-ball will start absorbing all anti-particles in his surroundings and grow. Then the Q-ball will evaporate
and vanish.


\begin{acknowledgments}
The author would like to thank M. Shaposhnikov for suggesting this problem, and M Ruiz-Altaba for discussions.
\end{acknowledgments}

\end{document}